\documentclass[final]{svjour2}
\usepackage{graphicx}
\usepackage{rotating}
\usepackage{amssymb}
\usepackage{mathptmx}
\usepackage[numbers]{natbib}

\makeatletter
\journalname{Journal of Low Temperature Physics}
\bibpunct{}{}{,}{s}{}{,}

\begin{document}

\title{Energy spectrum of the 3D velocity field, induced by vortex tangle}
\author{Sergey K.Nemirovskii}
\institute{Institute of Thermophysics\\Novosibirsk, 630090, Russia\\
\email{nemir@itp.nsc.ru}}
\date{XX.XX.2007}
\maketitle

\begin{abstract}
A review of various exactly solvable models on the determination
of the energy spectra $E (k) $ of 3D-velocity field, induced by
chaotic vortex lines is proposed. This problem is closely related
to the sacramental question whether a chaotic set of vortex
filaments can mimic the real hydrodynamic turbulence. The quantity
$<\mathbf{v(k)v(-k)}>$ can be exactly
calculated, provided that we know the probability distribution functional $%
\mathcal{P}(\{\mathbf{s}(\xi ,t)\})$ of vortex loops
configurations. The knowledge of $\mathcal{P}(\{\mathbf{s}(\xi
,t)\})$ is identical to the full solution of the problem of
quantum turbulence and, in general, $\mathcal{P}$ is unknown. In
the paper we discuss several models allowing to evaluate spectra
in the explicit form. This cases include standard vortex
configurations such as a straight line, vortex array and ring.
Independent chaotic loops of various fractal dimension as well as
interacting loops in the thermodynamic equilibrium also permit an
analytical solution. We also describe the method of an obtaining
the 3D velocity spectrum induced by the straight line perturbed
with chaotic 1D Kelvin waves on it.\newline

PACS number(s): 67.25.dk, 47.37.+q, 03.75.Kk
\end{abstract}

\section{Introduction}

One of the exciting applications of quantum turbulence is the solution (or
rather, the attempt at a solution) of the tantalizing problem of classical
turbulence. For this reason, the problem of the spectrum of the 3D velocity
field induced by a chaotic vortex filament becomes one of the central
questions. The formal relation, allowing of calculating $E(\mathbf{k})={\rho
}_{s}\left\langle \mathbf{v}_{\mathbf{k}}\mathbf{v}_{-\mathbf{k}%
}\right\rangle $ via the vortex line configuration $\{\mathbf{s}(\xi )\}$,
is given by formula (see \cite{Nemirovskii1998})
\begin{equation}
E=\left\langle \frac{{\rho }_{s}{\kappa }^{2}}{2}\int\limits_{\mathbf{k}}%
\frac{d^{3}\mathbf{k}}{\mathbf{k}^{2}}\int\limits_{0}^{L}\int\limits_{0}^{L}%
\mathbf{s}^{\prime }(\xi _{1})\mathbf{s}^{\prime }(\xi _{2})d\xi _{1}d\xi
_{2}\exp \left[ i\int\limits_{\xi _{1}}^{\xi _{2}}\mathbf{ks}^{\prime }(%
\tilde{\xi})d\tilde{\xi}\right] \right\rangle .  \label{E}
\end{equation}
In isotropic case, the spectral density depends on the absolute
value of
the wave number $k$. Integrating over solid angle leads to formula (see \cite%
{Kondaurova2005}):
\begin{equation}
E(k)=\left\langle\frac{\rho _{s}\kappa ^{2}}{(2\pi )^{2}}\int\limits_{0}^{L}\int%
\limits_{0}^{L}\mathbf{s}^{\prime }(\xi _{1})\mathbf{s}^{\prime
}(\xi
_{2})d\xi _{1}d\xi _{2}\frac{\sin (k\left\vert \mathbf{s}(\xi _{1})-\mathbf{s%
}(\xi _{2})\right\vert )}{k\left\vert \mathbf{s}(\xi
_{1})-\mathbf{s}(\xi _{2})\right\vert }\right\rangle.  \label{E(k)
spherical single}
\end{equation}%
For anisotropic situations, formula (\ref{E(k) spherical single})
is understood as an angle average. Further we will apply these
formulas to study some particular situation.\\
     Of course, the
relations (\ref{E(k) spherical single})-(\ref{E}) are just
mathematical identities, and the physics is hidden behind the
$\left\langle {}\right\rangle $ operation. To calculate
this average, we need the probability distribution functional $\mathcal{P}(\{%
\mathbf{s}(\xi )\},t)$, which is the\ probability that the system
has the configuration $\{\mathbf{s}(\xi )\}$ of the set of the
vortex loops. Knowing $\mathcal{P}(\{\mathbf{s}(\xi )\},t)$ is
identical to the full solution of the problem of quantum
turbulence. Therefore, the problem of the determination of the 3D
energy spectrum is not resolved, although there are a series of
theoretical approaches and numerical simulations. We consider here
several exactly solvable cases, the study of them is quite
instructive.
\section{Regular structures (straight line, vortex array, ring)}

The angle averaged spectrum created by straight vortex line
(directed along axis $z$) is directly evaluated from (\ref{E(k)
spherical single})

\begin{equation}
\int\limits_{\mathbf{k}}dk\frac{{\rho }_{s}{\kappa }^{2}}{(2\pi )^{2}}%
\left\langle \int\limits_{0}^{L}\int\limits_{0}^{L}\frac{\sin
(k(z_{1}-z_{2}))}{(k(z_{1}-z_{2}))}\;dz_{1}dz_{2}\right\rangle =\frac{{\rho }%
_{s}{\kappa }^{2}L}{4\pi }\int\limits_{\mathbf{k}}dk\frac{1}{k}.
\label{spectrum straight line}
\end{equation}%
The length $L$ is the conditional quantity. more correctly to talk here
about energy per unit length. Spectrum of the straight line (\ref{spectrum
straight line}) is discussed early (see, e.g., \cite{Nore1997},\cite%
{Nore1997a},), Vinen \cite{Vinen2000} proposed the $k^{-1}$\ spectrum on the
basis of dimensional consideration.\ This result is important, since it (\ref%
{spectrum straight line}) states that for any vortex system the high wave
numbers larger than inverse curvature, $E(k)$\ should be ($\rho _{s}\kappa
^{2}/4\pi )k^{-1}$ (per unit length).

Let us take a set of straight vortex filaments forming the square lattice $%
\bigcup s_{i}(\xi )=\bigcup (x_{i},y_{i},z)$. Points $x_{i},y_{i}$\ are
coordinates for vortices on the $xy$-plane, index $i$\ runs from $1$\ to $N$%
\ . The neighboring lines are separated by distance $b$, i.e., $%
 x_{i+1}-x_{i}=b$. Then the general relation (\ref{spectrum
straight line}) leads to the following formula (Hanninen (private
communication), Nowak et al. (\cite{Nowak2012}):
\begin{equation}
\frac{E(k)}{{\rho }_{s}{\kappa }^{2}L}=\frac{1}{4\pi k}\sum_{i=1}^{N}%
\sum_{i=1}^{N}J_{0}(kd_{ij})  \label{Risto lattice}
\end{equation}%
where $d_{ij}=\sqrt{(x_{i}-x_{j})^{2}+(y_{i}-y_{j})^{2}}$\ distances between
vortices on the $xy$-plane. Thus, determination of the spectrum on the basis
(\ref{Risto lattice}) should be done with the use of the quadruple summation
(over $(x_{i},x_{j},y_{i},y_{j})$), which requires large computing
resources. Clear, however, that for very small $k$, which corresponds to
very large distance, the whole array can be considered as large single
vortex with the circulation $N^{2}\kappa $. Accordingly, the spectrum (per
unit height) should be ($\rho _{s}N^{4}\kappa ^{2}/4\pi )k^{-1}$. For large $%
k$, which corresponds to very small distance from each line, the spectrum
(per unit height) should be $(\rho _{s}\kappa ^{2}/4\pi )k^{-1}$ In the
intermediate region $kb<<1$, and $Nkb>>1$\ (this condition implies that
inverse wave number $k^{-1}$\ is larger intervortex space between
neighboring lines, but smaller then the size of the whole array $Nd_{ij}$),
we can replace the quadruple summation by the quadruple integration with
infinite limits. This procedure corresponds that we exclude the fine-scale
motion near each of vortex, and are interested in the only large-scale,
coarse-grained motion. After obvious change of variables $x_{i}\rightarrow
kx_{i},\ y_{i}\rightarrow ky_{i}$\ etc. we get that the the whole integral
should scale as $1/k^{4}$, and accordingly
\begin{equation}
E(k)\varpropto 1/k^{5}  \label{minus 5}
\end{equation}

Formula $E(k)=dE/dk\varpropto 1/k^{5}$\ implies that
$E(\mathbf{k})=dE/d^{2}k$\ should behave as $1/k^{-6}$. It
follows, for example, from that in the 2D case $d^{2}k=2\pi kdk$\
(isotropic is assumed). Since, further,
$E(\mathbf{k})=v(\mathbf{k})v(-\mathbf{k})$, we see that
$v(\mathbf{k})$\ scales as $1/k^{-3}$. The latter means (see,
e.g., \cite{Frisch1995}, Eqs. (4.60),(4.61)) that the velocity
$v(\mathbf{r})$\ scales as $\mathbf{r}^{1}$. Thus, the uniform
vortex array creates the course-grained motion, which \ is
rotation, as it should be. Moreover, the coefficient is equal to
$\kappa /2b^{2}$, which coincides with the Feynman rule.

Now we will consider a vortex ring with radius $R$\ lying in the $x$-$y$\
plane. The line $s(\xi )$\ can be parameterized as $\mathbf{s}(\mathbf{x}%
)=\left( R\cos \varphi ,R\sin \varphi ,0\right) $ with $\varphi
\,\in \,[0,2\pi ]$. Applying it to Eq. (\ref{E(k) spherical
single}) we get (see also \cite{Nowak2012})

\begin{equation}
E_{ring}=\frac{\rho _{s}\kappa ^{2}R}{(2\pi )^{2}}\int\limits_{0}^{2\pi
}\int\limits_{0}^{2\pi }d\varphi _{1}d\varphi _{2}\frac{\cos (\varphi
_{1}-\varphi _{2})\sin (2kR\sin ((\varphi _{1}-\varphi _{2})/2))}{2k\sin
((\varphi _{1}-\varphi _{2})/2)}
\end{equation}

Evaluating the integral numerically shows that the spectrum
$E_{ring}(k)$\ scales like $E_{ring}(k)$ $\sim k^{2}$\ for $kR\ll
1$. Frequently, the $k^{2}$ is referred to as a proof for the
thermodynamical equilibrium state. We would like to stress,
however, that this distribution of the energy (valid far from the
ring) has nothing to do with the equipartition law. It is a
consequence of the fact that closed vortex loops induce a far
field flow scaling as $1/r^{3}$. That, in turn, generates a
spectrum $E(k)\propto
k^{2} $. This fact was established for classical turbulence (see, e.g., \cite%
{Monin1975}). For quantum turbulence this result was discussed by Stalp,
Skrbek and Donnelly \cite{Stalp1999}. For the large $k$, namely for $kR\gg 1$%
, spectrum $E_{ring}(k)$\ scales like $\sim k^{-1}$ as for
straight line.

\section{Noninteracting Gaussian loops of various fractal dimension}

One of the approaches, allowing exact solution, based on the
viewing of the VT as a set of loops having the random structure
with the various fractal dimension \cite{Nemirovskii2002}. This
theory is based on the Gaussian model (see
\cite{Nemirovskii1998}). The probability
$\mathcal{P}_{Gauss}(\left\{
\mathbf{s}(\xi ,t)\right\} )$\ of finding a particular configuration $%
\left\{ \mathbf{s}(\xi ,t)\right\} $ is expressed by the probability
distribution functional (for details, see the paper by author \cite%
{Nemirovskii1998})
\begin{equation}
\mathcal{P}_{Gauss}(\left\{ \mathbf{s}(\xi ,t)\right\} )=\mathcal{N}\exp
\left( -\int\limits_{0}^{l}\int\limits_{0}^{l}\mathbf{s}^{\prime \alpha
}(\xi _{1},t)\Lambda _{\alpha \beta }(\xi _{1}-\xi _{2})\mathbf{s}^{\prime
\beta }(\xi _{2},t)d\xi _{1}d\xi _{2}\right) .  \label{Gauss_model}
\end{equation}%
Here $\mathcal{N}$ is a normalizing factor and $l$ is the length
of curve. The typical form of function $\Lambda _{\alpha \beta
}(\xi -\xi ^{\prime })$ is a smoothed $\delta $ function, having a
Mexican hat shape\ with width equal to $\xi _{0}$. Behavior of
$\Lambda _{\alpha \beta }(\xi _{1}-\xi _{2})$ on large scales
determines the second order correlation function between
tangent vectors $\langle \mathbf{s}_{\alpha }^{\prime }(\xi _{1})\mathbf{s}%
_{\alpha }^{\prime }(\xi _{2})\rangle $, which, in turn, determines the
fractal dimension of line. Referring the reader to the original article \cite%
{Nemirovskii2002} for the details we briefly state the main results.

\begin{figure}[tbp]
\includegraphics[width=12cm]{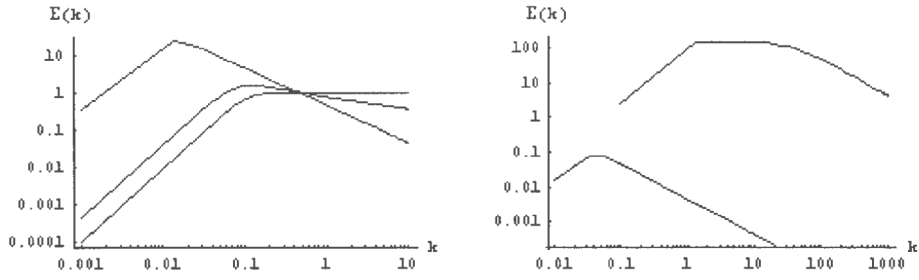}
\caption{(Color online) Energy spectra of a pure fractal line (left), and a
more realistic semi-fractal vortex loop (right).}
\label{fractalspectra}
\end{figure}

In the region $k\ll 1/L^{1/H_{D}}$, the spectrum behaves as $E(k)\propto
k^{2}$. We again stress that this dependence is related to the far field
flow $1/r^{3}$ from the restricted domain of vorticity (but not to the
thermodynamical equilibrium). Note, that $L^{1/H_{D}}$ is nothing but the
real 3D size of the loop. In the opposite case, of large $k$ we get
\begin{equation}
E(k)\propto k^{-2+H_{D}}.  \label{E fractal}
\end{equation}%
This result had been obtained earlier from qualitative
considerations and is discussed in Frisch \cite{Frisch1995}. Fig.
\ref{fractalspectra}
(left) depicts (in logarithmic scale) three curves, the spectral densities $%
E(k)$ for a pure fractal vortex filament of length $L=100$ (all units are
arbitrary) and three different values of $H_{D}=1,5/3,2$ correspondingly. It
can be seen how $E(k)\propto k^{2}$ is changed to a dependence $E(k)\propto
k^{-2+H_{D}}$ in the region of large $k\sim 1/100^{-1/H_{D}}$.

Vortex loops in superfluid turbulent quantum fluids are "semifractal"
objects, which are smooth at small $(\xi ^{\prime }\;-\;\xi ^{\prime \prime
})$ along the curve, and are fully uncorrelated for the remote parts. A good
approximation for $\langle \mathbf{s}_{\alpha }^{\prime }(\xi _{1})\mathbf{s}%
_{\alpha }^{\prime }(\xi _{2})\rangle $ is a function of the type $%
1/(1+(\Delta \xi /\xi _{0})^{2})$.

Fig. \ref{fractalspectra} (right) shows (in logarithmic scale) the
spectral densities $E(k)$ for a "semifractal" vortex loop of
length $L=100$ with $\xi _{0}=0.01$ and $\xi _{0}=1$. It is
clearly seen that the curves ($\xi _{0}=0.01$) have three regions
with bends at the points $k_{l}=1/\sqrt{L\xi _{0}}=1$, and
$k_{r}=1/\xi _{0}=100.$ In these three different regions we have
$E(k)\propto k^{2}$, $E(k)\propto k^{0}$ and $E(k)\propto k^{-1}$,
correspondingly. This is in excellent agreement with the
qualitative considerations.

\section{Thermodynamically equilibrium vortex tangle}

\label{spectrum interacting}

In the previous Section, we considered the model of free
(noninteracting) vortex loops. Let us now discuss another solvable
case, that of interacting loops in thermodynamical equilibrium.
This can be accomplished by
introducing the Boltzmann factor $\exp (-\beta H\left\{ \mathbf{s}\right\} )$%
into the Gaussian model (\ref{Gauss_model}) (see
\cite{Nemirovskii2004,Nemirovskii2009}). Then the partition
function can be written (we will take here the purely Wiener
distribution with $\Lambda$ being the $\delta$ function and with
an elementary step equal to $\xi _{0}$)

\begin{equation}
Z=\int D\mathbf{s}(\xi \}\exp \left[ -\frac{3}{2\xi _{0}}\int\limits_{0}^{L}(%
\mathbf{s}^{\prime }(\xi ))^{2}d\xi -\beta \frac{{\rho }_{s}{\kappa }^{2}}{%
8\pi }\oint \oint d\xi _{1}d\xi _{2}\frac{\mathbf{s}^{\prime }(\xi _{1})%
\mathbf{s}^{\prime }(\xi _{2})}{\left\vert \mathbf{s}(\xi _{1})-\mathbf{s}%
(\xi _{2})\right\vert ^{3}}d\xi _{1}d\xi _{2}\right]  \label{Z initial}
\end{equation}

Here $\beta =1/k_{B}T$ \ is the inverse temperature. A
considerable simplification in the evaluation of the partition
function can be reached with the use of the Edwards trick
(see for details \cite{Kleinert1990},\cite{Edwards1979}, \cite{Copeland1991},%
\cite{Nemirovskii2012}), namely, $\exp (-\beta H\left\{ \mathbf{s}\right\} )$
can be written as a the Gaussian path integral over an auxiliary vector
field $\mathbf{A}(\mathbf{r)}$. After all the transformations, the partition
function acquires the form of a Gaussian path integral over the 3D auxiliary
vector field $\mathbf{A}(\mathbf{r)}$ and $E=d\ln Z/d\beta $ is (see \cite%
{Nemirovskii2012}):%
\begin{equation}
E=\frac{3}{4\pi ^{2}}\xi _{0}L{\rho }_{s}{\kappa }^{2}\int dk\frac{k^{2}}{%
(k^{2}+M^{2})}
\end{equation}%
with $M^{2}=\beta \xi _{0}L{\rho }_{s}{\kappa }^{2}/2\mathcal{V}$
($\mathcal{V}$ is the volume). For small $k$, the energy spectrum
$E(k)\varpropto (k_{B}T)k^{2}$, which is nothing but the Rayleigh
equipartition law $dE/d^{3}\mathbf{k=}const$. For large
wave numbers $E(k)$ is just about constant. The reason is that for large $k$%
, the interaction energy in the partition function (\ref{Z initial}) is
smaller and the main contribution appears from the configuration term $(%
\mathbf{s}^{\prime }(\xi ))^{2}$, which is related to the
connectivity of the line. Hence, we have the same spectrum as for
a pure random walk with $H_{D}=2$, discussed in previous Section.

\section{1D Kelvin waves spectrum and 3D velocity spectrum}

In the literature there is discussed the idea of obtaining the 3D
velocity spectrum just by putting it equal to the spectrum of 1D
Kelvin waves. For instance, as stated in \cite{Kivotides2001b}:"We
notice that, because the fluctuations of the velocity field are
induced by the Kelvin wave fluctuations on the filaments, it is
reasonable \ to expect that
\begin{equation}
E(k)\sim E_{KW}(k)".  \label{Ev equal Ehd}
\end{equation}%
The same idea was used in papers by L'vov et al.(see e.g., \cite{Lvov2008}).
Details of this activity can be read in a series of papers by L'vov,
Nazarenko and coauthors \cite{L'vov2007}.\cite{Lvov2008,Sasa2011,Lvov2011}

Let us consider this problem on the basis of general formula
(\ref{E}) (see \cite{Nemirovskii2012}). We take $\mathbf{s}(\xi
,t)=(x(z,t),y(z,t),z)$ and denote the two-dimensional vector
$(x(z,t),y(z,t))$\ as $a\rho (z,t)$\
(where the dimensionless amplitude $a\ll 1$). Substituting it into (\ref%
{E(k) spherical single}) and expanding in powers of $a$, we get,
\begin{eqnarray}
E &=&E_{0}+a^{2}\frac{{\rho }_{s}{\kappa }^{2}}{2}\int\limits_{\mathbf{k}}%
\frac{d^{3}\mathbf{k}}{\mathbf{k}^{2}}\int\limits_{0}^{L}\int%
\limits_{0}^{L}dz_{1}dz_{2}{\LARGE \{}\frac{\cos \left( k\left\vert
z_{2}-z_{1}\right\vert \right) (\mathbf{\rho }(z_{2})-\mathbf{\rho }%
(z_{1}))^{2}}{2\left\vert z_{2}-z_{1}\right\vert ^{2}}  \label{E(k) with KW}
\\
&&-\frac{\sin \left( k\left\vert z_{2}-z_{1}\right\vert \right) (\mathbf{%
\rho }(z_{2})-\mathbf{\rho }(z_{1}))^{2}}{2k\left\vert
z_{2}-z_{1}\right\vert ^{3}}+\frac{(\mathbf{\rho }^{\prime }(z_{1})\mathbf{%
\rho }^{\prime }(z_{2})\sin \left( k\left\vert z_{2}-z_{1}\right\vert
\right) }{k\left\vert z_{2}-z_{1}\right\vert }{\LARGE \}}
\end{eqnarray}%
The first term $E_{0}$ of the zero-order in amplitude $a$\ exactly coincides
with the energy spectrum induced by the unperturbed straight line (\ref%
{spectrum straight line}), as it should be.

To move further we have to find the correlation characteristics for the
fluctuating vector of displacement $\rho (z_{2})$. We accept that the
ensemble of Kelvin waves has a following power-like spectrum:

\begin{equation}
\left\langle \mathbf{\rho }(p)\mathbf{\rho }(-p)\right\rangle =Ap^{-s}.
\label{KW- E(k) spectrum}
\end{equation}%
We take here the notation $p$\ for the one-dimensional vector,
conjugated to $z$, reserving the notation $k$\ for the absolute
value of the wave vector of the 3D field. The formula (\ref{KW-
E(k) spectrum}) implies that (see, e.g., \cite{Frisch1995}, Eqs.
(4.60),(4.61)) the squared increment for the vector of
displacement scales as, $\left\langle (\mathbf{\rho
}(z_{2})-\mathbf{\rho }(z_{1}))^{2}\right\rangle \varpropto
(z_{2}-z_{1})^{s-1}$. Then the second
order correlator $\left\langle (\mathbf{\rho }^{\prime }(z_{2})\mathbf{\rho }%
^{\prime }(z_{1}))\right\rangle $\ scales as $\left\langle (\mathbf{\rho }%
^{\prime }(z_{2})\mathbf{\rho }^{\prime }(z_{1}))\right\rangle \varpropto
(z_{2}-z_{1})^{s-3}$. Substituting it into (\ref{KW- E(k) spectrum}) and
counting the powers of quantity $k$, we conclude that the correction $\delta
E(k)$\ to the spectrum $E(k)$,\ due to the ensemble of Kelvin waves has a
form:%
\begin{equation}
\delta E(k)\varpropto {a}^{2}k^{-s+2}.  \label{dE due to KW}
\end{equation}%
It is remarkable fact that this quantity coincides formally with
the one-dimensional spectrum of KW $\delta E(p)\varpropto
a^{2}p^{-s+2}$\ however this contribution is small, by virtue the
smallness of the wave amplitudes $a$, and disappears with the KW.

\section{Conclusions}

Summarizing, it can be concluded that the 3D energy spectrum $E(k)$ consists
of several parts. At small $k$, associated with the large scales, on the
order of the size of the system, or on the scale of the stirring forcing
(grids, propellers, vibrating objects, etc.), the spectrum behaves as $%
E(k)\varpropto $ $k^{2}$. We recall again that this is the
consequence of the asymptotic behavior ($r\rightarrow \infty $) of
the velocity field, not of thermodynamical equilibrium. for large
wave numbers, exceeding the inverse intervortex space $k>2\pi
/\delta $, the energy spectrum $E(k)$ should be close to $k^{-1}$,
again regardless of the specific model. The region of intermediate
$k$ is the most intriguing and exciting, it depends on fractal
properties of lines. Unfortunately, so far there is no theory,
predicting the Kolmogorov like spectrum $E(k)\varpropto $
$k^{-5/3}$.

\begin{acknowledgements}
The work was supported by the grants N 10-08-00369 and N 10-02-00514 from
the Russian Foundation of Basic Research, and by the grant from the
President Federation on the State Support of Leading Scientific Schools
NSh-6686.2012.8.

\end{acknowledgements}

\pagebreak

\end{document}